\documentclass[12pt]{iopart}
\usepackage{iopams} 
\usepackage{latexsym}
\usepackage{epsf}
\newcommand{\C}{{\mathbb C\hspace{0.05 ex}}}
\newcommand{\tihop}{\ope{\rho}}
\newcommand{\tihep}{\tihop_\varepsilon}
\newcommand{\ham}{\ope{H}}
\newcommand{\norm}[1]{\Vert #1\Vert}
\newcommand{\apriori}{{\em a priori\/}}
\newcommand{\gmean}[1]{\mean{#1}^{\rm gauss}}
\newcommand{\be}{E'}
\newcommand{\bep}{\vep'}
\newcommand{\barh}{\overline{H}}
\newcommand{\trp}[1]{{#1}^{\rm T}}
\newcommand{\opm}{\ope{P}_{\pm}}
\newcommand{\A}{\ope{A}}
\newcommand{\T}{\ope{T}}
\newcommand{\Zcl}{Z_{\rm cl}}
\newcommand{\vep}{\varepsilon}
\newcommand{\mean}[1]{\langle #1\rangle}
\newcommand{\half}{{1\over 2}}
\newcommand{\re}{{\rm Re\,}}
\newcommand{\im}{{\rm Im\,}}
\newcommand{\ope}[1]{\widehat{#1}}
\newcommand{\set}[1]{\{#1\}}
\newcommand{\binom}[2]{{\biggl(
 \begin{array}{@{\,}c@{\,}} #1 \\[-2 pt] #2 \end{array}\biggr)}}

\newtheorem{theorem}{Theorem}
\newtheorem{proposition}[theorem]{Proposition}
\newtheorem{lemma}[theorem]{Lemma}

\newenvironment{proof}{\begin{trivlist}\item[]{\em %
 Proof:}\/}{\hfill\mbox{$\Box$}\end{trivlist}}

\begin{document} 

%\preprint{HIP-2000-03/TH}

% Journal identifier can be put here if required, e.g.
\jl{1}

\title[Lattice computation of energy moments]{%
 Lattice computation of energy moments in canonical and Gaussian
 quantum statistics} 
\author{Jani Lukkarinen\footnote[1]{E-mail address:
 {\tt jani.lukkarinen@helsinki.fi} }}
\address{Helsinki Institute of Physics, PO Box 9, 
 00014 University of Helsinki, Finland} 

\begin{abstract}
We derive a lattice approximation for a class of equilibrium quantum
statistics describing the behaviour of any combination and number of
bosonic and fermionic particles with any sufficiently binding
potential. We then develop an intuitive Monte Carlo algorithm which
can be used for the computation of expectation values in canonical and
Gaussian ensembles and give lattice observables which will converge to
the energy moments in the continuum limit.  The focus of the
discussion is two-fold: in the rigorous treatment of the continuum
limit and in the physical meaning of the lattice approximation.  In
particular, it is shown how the concepts and intuition of classical
physics can be applied in this sort of computation of quantum effects.
We illustrate the use of the Monte Carlo methods by computing
canonical energy moments and the Gaussian density of states for
charged particles in a quadratic potential.
\end{abstract}

\pacs{05.30.-d, 02.50.Ng, 05.10.Ln}

\submitted
%\maketitle

\section{Introduction}

Recently, the role of the canonical Boltzmann-Gibbs ensemble as the
only useful ensemble for quantum statistics has been reevaluated.  The
need for reevaluation has arisen both from experimental and
theoretical considerations.  Although the canonical ensemble works
surprisingly well also for systems consisting of only a few particles,
relativistic ion collisions have provided an arena where the
corrections from the finite size of the system to the canonical
results have become important \cite{redlich:94}.  A
quantitative understanding  of these collisions is necessary since
they are central in the search for new phases of strongly interacting
matter, such as quark-gluon plasma.  In addition, methods for a
quantitative measurement of the properties  of small quantum systems
could be useful in industrial applications in the future,  such as in
the design of ever smaller electronic components.  Similarly, the
functioning of certain biological systems  seems to involve also the
quantum behaviour of the system---consider, for instance, the
recognition of molecules by olfactory receptors
\cite{wright:smell}.

The second motivation for non-canonical ensembles comes from the
peculiar behaviour associated with systems with fractal-like
structure.  Tsallis statistics \cite{tsa88} was
developed for the statistical analysis of these cases and we have
considered other generalized statistics of that kind in a previous
work \cite{jml:tsallis}.  For ``ordinary'' systems, however, we found
out that the so called Gaussian ensemble is the simplest
generalization of the canonical ensemble which could be used for the
approximation of any precanonical system---these are systems for which
the occupation of high-energy states decreases faster than
exponentially.  For a more complete description of precanonical and
Gaussian ensembles, see \cite{jml:tsallis} and  \cite{jml:gauss},
respectively.  Of the non-canonical ensembles, we concentrate here
only on the Gaussian ensemble with the understanding that  the results
presented can be easily modified to apply  for any precanonical
ensemble.

In this work, we develop and prove the existence of a lattice
approximation for a system of a fixed number of non-relativistic
particles which may be fermions, bosons or a combination of both.  The
potential of this system can be very general, the only requirement
being a fast enough increase at infinity.  We do not consider the
possibility of a creation or annihilation of particles---this would
require the use of a quantum field theory instead of ordinary quantum
mechanics and we would have to give up mathematical rigour in the
transition.  Formulations of the microcanonical ensemble using quantum
field theory and its perturbative expansion have already been proposed
by Chaichian and Senda \cite{chs:93} and our approach owes a
lot to their work.

We begin by defining notations and by stating assumptions we need to
make on the potential in section \ref{sec:notations}.   We then state
a theorem which defines and proves the convergence of the lattice
approximation of traces at a complex temperature and  we derive a set
of lattice observables which can be used for the measurement of energy
moments. The following section \ref{sec:canalg} presents a Monte Carlo
algorithm  for the generation of the canonical lattice distribution
and we also try to develop an intuitive understanding of it.   Section
\ref{sec:gauss} contains similar derivations for the Gaussian
ensemble, especially it contains an essentially canonical  algorithm
which can be used for the computation of  the density of states.  We
conclude with an example where we test the algorithms in practice and
comment on their strengths and draw-backs.  We have also gathered a
comparison  of our methods to the ones presented in the literature
along with suggestions for improvements in section \ref{sec:prologue}.
A good review of the state of the art of using Monte Carlo simulations
for path-integrals of quantum fluids has been given by Ceperley
\cite{ceperley95}.

\section{Notations and background}\label{sec:notations}

Consider a system of $N$ particles living  in a $d$-dimensional space.
In classical physics, this system is described by trajectories of
these particles and the positions of the particles form the relevant
degrees of freedom, an $n$-dimensional space with $n=N d$.  We shall
here choose the convention that the positions of particles are denoted
by boldface letters with a subscript identifying the label of the
particle while the classical configuration of the system  is denoted
by regular italics. For example, the $n$-dimensional  configuration
$x$ could be written in terms of the $N$ position vectors as
$x=(\bi{x}_1,\ldots,\bi{x}_N)$.

The lattice regularization of the quantum effects on this system is
defined on a space containing $L$ copies of the classical
configuration space.  The ``lattice configurations'' are thus  
$n L$-dimensional vectors and we shall put the number of the copy 
in parenthesis after the configuration symbol.  In other words, the
position of the particle $i$ in the copy $k$ in the lattice
configuration $x$ is denoted by $\bi{x}_i(k)$.  The quantum results
are recovered in the continuum limit, by which we mean here simply
taking  $L$ to infinity in the lattice regularized formulae given
below.

The following discussion applies to massive, non-relativistic
particles which are bounded by a suitably strong potential and thus
can be described by equilibrium ensembles.  To simplify matters, we
also assume that all masses have been scaled away by the
transformation $\bi{x}_i \mapsto \bi{x}_i/\sqrt{m_i}$ and we use the
natural units where the Planck and Boltzmann constants are equal to
one.

In summary, we assume that the Hamiltonian of the system is given by
\[
\ham=\sum_{i=1}^N \half \ope{\bi{p}_i}^2 +
 V(\ope{\bi{x}}_1,\ldots,\ope{\bi{x}}_N)
\]
and the precise conditions for the interaction potential are
\begin{enumerate}
\item\label{it:Vcon} $V$ is continuous and bounded from below.
\item\label{it:potcon}  $\int \rmd^n y\, \rme^{-\beta V(y)} <\infty$
  for all $\beta>0$.
\item\label{it:Vperm}  $V$ is invariant under permutations of
  indistinguishable particles.
\end{enumerate}
With the first condition, the Hamiltonian can be defined in the sense
of distributions (Theorem X.32 in \cite{ReedSimonII}) and the
continuity also ensures that the continuum limit will be simple---if
the original potential is not continuous, it can be replaced by a
suitable continuous approximation.  The second condition requires a
sufficiently fast growth of the potential at infinity and it gives the
precise meaning for what we meant by a sufficiently binding potential
earlier.  The third requirement is more of a consistency condition
than a restriction.

The classical partition function of this system is defined by
\begin{equation}\label{e:defZcl}
\Zcl(\beta) \equiv  \int\!\! \frac{\rmd^n y}{(2\pi\beta)^{n/2}}\,
  \rme^{-\beta V(y)}
\end{equation}
and clearly (\ref{it:potcon}) is equivalent to the assumption that the
classical partition function converges for all positive temperatures.
It is also clear that if $V$ is positive, which could be achieved by
adding a constant to the potential, then $\Zcl(\beta)$ decreases
monotonically.

For the sake of simplicity, we shall now assume that there is only one
kind of indistinguishable particles in the system.  The generalization
of the results to systems with many types of particles should be
obvious.

For each element $s$ in the permutation group $S_N$  we can define a
linear operator which performs the corresponding particle permutation
by 
$s(\bi{x}_1,\ldots,\bi{x}_N)= (\bi{x}_{s(1)},\ldots,\bi{x}_{s(N)})$.  
These operators form a unitary
representation of the permutation group, i.e.\ all operators are
orthogonal, $\trp{s}s=1$.  The determinant of an operator will also be
equal to the parity of the corresponding permutation,  denoted here by
$(-)^s$.  With these notations, the  projection operator $\opm$ on the
physical Hilbert space can be defined by
\[
\opm\psi(x) = \frac{1}{N!} \sum_{s\in S_N} (\pm )^s \psi(sx)
\]
where $(+)^s = +1$ and the upper sign is used for bosons and  the
lower for fermions.   The trace of any trace-class operator $\A$  over
the physical subspace can also be obtained from
\[
\Tr_{\rm phys}\!\left(\A\right) = \Tr\!\left(\opm\A\right) =
\Tr\!\left(\A\opm\right).
\]
Note also that condition (\ref{it:Vperm}) is equivalent to requiring
that $V(sx)=V(x)$ for all permutations $s$.

\section{Definition and convergence of the lattice approximation}
\label{sec:deflat}

We shall now state a theorem which can be used for getting lattice
observables for energy moments and which proves the convergence of the
continuum limit of the lattice approximation.  We prove the theorem
for the canonical case with a temperature that can have an imaginary
part---this complex temperature trace is needed for the computation of
the energy moments in precanonical ensembles as we shall see later.
The proof of the theorem is given in \ref{sec:latproof}.
\begin{theorem}\label{th:latconv}
Define 
$\ope{T}(z)= \exp({-z\frac{1}{2} \ope{p}^2})\exp({-z V(\ope{x})})$ 
on the half-plane $\re z > 0$.  Then for all $k\ge 0$ and $\re z >0$,
\[
\Tr\!\left(\opm\ham^k\e^{-z\ham}\right) =  \lim_{L\to\infty}
 (-1)^k\frac{\rmd^k}{\rmd z^k} \Tr\!\left(\opm \ope{T}(z/L)^L\right).
\]
All of these functions are analytic on the right half-plane and the
convergence is uniform on compact subsets of the half-plane.  In
addition, for any fixed $k$ and $0<c\le c'$, these functions are
uniformly bounded on the closed strip  $c\le\re z\le c'$ of the
half-plane.
\end{theorem}

The proof contains the following explicit representation for  the
traces on the right hand side
\begin{equation}\label{e:trppm}
\Tr\!\left(\opm \ope{T}(z/L)^L\right) =  \frac{1}{N!} 
  \sum_{s\in S_N} (\pm )^s \int\!\rmd^{nL} x\, K_L(sx(L),x;z),
\end{equation}
where the ``lattice kernel function'' is defined by
\[
K_L(b,x;z) = \biggl( {L\over 2 \pi z} \biggr)^{{L n\over 2}}
  \exp\!\bigg[ -\frac{1}{z} P_L(b,x) - z V_L(x)\bigg]
\]
and the ``lattice kinetic energy'' $P_L$ and the ``lattice potential
energy'' $V_L$ are
\begin{eqnarray*}
P_L(b,x) = {L\over 2}\sum_{k=1}^L |x(k)-x(k-1)|^2 
  \mbox{, with }x(0)=b, \\ 
V_L(x) = {1\over L}\sum_{k=1}^L V(x(k)).
\end{eqnarray*}
This result is well-known and widely used,   it was presented already
in the classic work by Feynman and Hibbs \cite{FHibbs}.  However, to
our knowledge it has not been rigorously proven before  that the
lattice approximation of the complex temperature trace does not spoil
any part of the result.  Note that theorem \ref{th:latconv} is not a
simple consequence of the Trotter product formulae
\cite{trotter59} as these imply only strong convergence  for
our generators: we have to do more work for proving the convergence of
the traces.

The sum in (\ref{e:trppm}) contains $N!$ terms and thus its
computation would be impossible for even fairly small values of $N$.
However, the result of the integral in fact only depends on the
conjugate class of the permutation and thus the $N^N$ increase of the
number of terms to  compute can be reduced to an exponential $\rme^N$
increase.  To be precise, it is shown in \ref{sec:perms}, that
\begin{eqnarray}
\fl \Tr\!\left(\opm \ope{T}(z/L)^L\right) = \frac{1}{N!}
  \sum_{\lambda_1=1}^N (\pm 1)^{N-\lambda_1}
  \sum_{\lambda_2=0}^{\lambda_1}\cdots
  \sum_{\lambda_N = 0}^{\lambda_{N-1}}
  \delta\!\left(\sum_{l=1}^N\lambda_l - N\right) \nonumber\\ 
\times c_{(\lambda)} \int\!\rmd^{nL} x\, 
  K_L(s_{(\lambda)}x(L),x;z),\label{e:conjsum}
\end{eqnarray}
where $s_{(\lambda)}$ is a representative of the conjugate class
defined by the parameters $(\lambda_l)$ as  explained in
\ref{sec:perms} and  $c_{(\lambda)}$ counts the number of elements in
the conjugate class, as given by equation \ref{e:defcnu}.  The sum
goes over those decreasing sequences of $N$ non-negative integers
whose sum equals $N$; in the above equation, $\delta$  is the
Kronecker symbol which ensures that the last condition holds.

If all the terms in the sum are relevant, then only about  $N\le 30$
can be handled by the above formula (for $N=30$ there are $5604$ terms
in the sum).  However, typically only a fraction of the terms are
relevant and the sum can be used also for a larger number of
particles.  For real temperatures,  the estimation of which terms will
be relevant can be done fairly easily by using lemma \ref{th:ikbound}
given in \ref{sec:latproof}.  According to the lemma, for all
$\beta>0$,
\begin{equation}\label{e:classbound}
\int\!\rmd^{nL} x\, K_L(sx(L),x;\beta) \le \int\!\! 
  \frac{\rmd^n y}{(2\pi\beta)^{n/2}}\,
  \exp\!\bigg[{-\frac{1}{2\beta} |sy-y|^2-\beta V(y)}\bigg],
\end{equation}
where the right hand side is much easier to evaluate than the left
hand side.   This also gives one more proof of the well-known result
that only the identity permutation is relevant for dilute gases, for
which the typical classical distance between the particles is larger
than the thermal wavelength $\sqrt{2\pi\beta}$.

\section{Lattice operators for energy moments}\label{sec:latmom}

We shall next derive a set of lattice operators whose expectation
values will converge in the continuum limit to the canonical
expectation values of powers of the Hamiltonian.  By the results of
section \ref{sec:deflat},
\[
\Tr\!\left(\opm\ham^k\e^{-z\ham}\right) =  \lim_{L\to\infty}
 \frac{1}{N!} \sum_{s\in S_N} (\pm )^s \int\!\rmd^{nL} x\,
 (-1)^k\frac{\rmd^k}{\rmd z^k} K_L(sx(L),x;z),
\]
where we have moved the differentiation inside the integral---it is
part of the proof in \ref{sec:latproof} that  this is allowed.

By using the Leibniz rule and a bit of algebra we can compute the
derivatives,
\[
(-1)^k\frac{\rmd^k}{\rmd z^k} K_L(sx(L),x;z) =  Q_k(x; L,z)
  K_L(sx(L),x;z),
\]
where
\begin{equation}\label{e:defQk}
Q_k(x; L,z) = \sum_{j=0}^k \binom{k}{j} V_L^{k-j}  \sum_{i=0}^j
  \binom{j}{i}  \frac{(\case12 L n-1+j)!}{(\case12 L n-1+i)!}
  (-P_L)^i z^{-j-i} .
\end{equation}
We call $Q_k$ the lattice operator for the $k$:th energy moment.   The
complicated form might be a bit surprising---the lattice action, after
all, is simple and it can be guessed from   the expected continuum
limit.  We might thus expect to find a lattice Hamiltonian equally
easily, the guess ``$\beta H_L$''=$\beta V_L- \case{1}{\beta} P_L$
which goes to 
``$\int_0^\beta\rmd\tau [V(x(\tau))-\case12 \dot{x}^2(\tau)]$'' 
in the continuum limit, being most natural.

In fact, the natural guess is not too far away, since by
(\ref{e:defQk})
\begin{equation}\label{e:defQ1}
\beta Q_1(x;L,\beta) = \beta V_L - \case{1}{\beta} P_L + \case12 L n,
\end{equation}
measures the expectation value of the Hamiltonian normalized by the
inverse temperature $\beta$.  The important difference is that the
other energy moments are {\em not}\/ measured by $Q_1^k$. Instead,
each power needs a separate ``renormalization'' term, note that the
extra factors all contain $\case12 L n$ which diverges in the
continuum limit.

We have given (\ref{e:defQk}) in a form from  where it is apparent
that it is the ``kinetic energy'' part of the Hamiltonian, given by
$P_L$,  which gets renormalized.  This is very natural, since the path
integration which gives the continuum results goes over continuous but
non-differentiable paths and thus the derivatives should diverge in
the continuum limit.  In addition to (\ref{e:defQ1}) we have
\begin{equation}\label{e:defQ2}
Q_2(x;L,z) = Q_1^2(x;L,z) + \frac{1}{z^2}  
  \left( \case12 L n -2 P_L \right),
\end{equation}
which can be used for concluding that both the expectation value and
variance of the lattice operator $P_L$ diverge as  $\case12 L n z$ in
the continuum limit.

\section{Monte Carlo algorithm for the generation 
of the canonical lattice distribution}\label{sec:canalg}

We saw in section \ref{sec:latmom} that the term $P_L$ diverges
proportionally to the lattice size $L$ in the continuum limit.  On the
other hand, $V_L$ has a finite continuum limit.  Consider generating
the lattice kernel distribution 
$\propto \exp({-\beta V_L -\case1\beta P_L})$ 
by a straightforward Metropolis algorithm
\cite{momu:qfl},  i.e.\ make random, local, changes to the lattice
configuration and keep the changes with the probability 
$\exp(-\Delta S_\beta)$.  
Since $P_L$ diverges, for large $L$ most changes will be
rejected because of the large jump they cause in $P_L$ and the
hit-rate of the Metropolis algorithm is not very good.  But, on the
other hand, $P_L$ is a simple quadratic function which does not depend
on the system, i.e.\ on the potential or on the nature of the
particles.  We shall now present an algorithm which takes advantage of
these simplifications and which  behaves better for large $L$.

The aim is to generate the canonical lattice distribution for $x$,
\[
\int\!\rmd^{nL} x\, 
  \biggl( {L\over 2 \pi \beta} \biggr)^{{L n\over 2}} 
  \exp\!\bigg[ -\frac{1}{\beta} P_L(sx(L),x) - \beta V_L(x)\bigg],
\]
where $s$ is any permutation and $\beta>0$.  Let us first separate the
lattice copy containing the jump induced by the permutation and define
$y=x(L)$. In the remaining $n(L-1)$ integrals we make the following
change of variables from $x$ to $u$,
\begin{equation}\label{e:deful}
x(k) = c(k) + \sum_{l=1}^k u(l)\mbox{, for all }1\le k \le L-1,
\end{equation}
where $c(k)= (1-\case{k}{L}) sy + \case{k}{L} y$.  The Jacobian of
this change of variables is one.  As earlier, for $k=0$ and $k=L$ we
define also  $x(0)=sy$ and $x(L)=y$.  The inverse relation is then
$u(k)=x(k)-x(k-1)-\case1L (y-sy)$ which shows that
\begin{eqnarray*}
\fl P_L(sx(L),x) = {L\over 2}\sum_{k=1}^L |x(k)-x(k-1)|^2 \\ 
\lo= {L\over 2}\sum_{k=1}^{L-1} |u(k)|^2 + 
 {L\over 2}\bigg|\sum_{k=1}^{L-1} u(k)\bigg|^2 
 + {1\over 2} |y-sy|^2.
\end{eqnarray*}
A choice of other than a straight path $c(k)$ does not lead to as
simple a formula, since then there would be a cross term linking $y$
and $u$ which is absent for a path with constant increments of $c$.

For reasons that will become apparent in the next section, we shall
use a separate notation $\beta'$ for the ``kinetic temperature''.
Thus with the assumption that at this point $\beta'=\beta$, we can
express the lattice integral as \numparts
\begin{eqnarray}\label{e:latalg}
\fl \int\!\! \frac{\rmd^n y}{(2\pi\beta')^{n/2}}\,
  \exp\!\bigg[{-\frac{1}{2\beta'} 
    |sy-y|^2-\beta V(y)}\bigg]\label{e:genclass}\\ 
\times \int\!\!\frac{\rmd^{n(L-1)} v}{(2 \pi)^{(L-1) n/2}} 
  \,\rme^{-\frac{1}{2} v^2} L^{\frac{n}{2}}
  \exp\!\bigg[-\frac{1}{2} \sum_{i=1}^N  \bigg|\sum_{k=1}^{L-1}
  \bi{v}_i(k)\bigg|^2 \bigg]  \label{e:genqf}\\ 
\times \exp\!\bigg\{-\frac{\beta}{L}
  \sum_{k=1}^{L-1} [V(x(k))-V(y)]\bigg\},\label{e:genrest} 
\end{eqnarray}
\endnumparts where we have scaled the temperature away from the
lattice fluctuations, i.e.\ made the change of variables
$u=\sqrt{\beta'/L}\, v$. This means that in the last line we have used
the definition
\begin{equation}\label{e:deftl}
x(k) = sy + \frac{k}{L} (y-sy) + \sqrt{\frac{\beta'}{L}}  
  \sum_{l=1}^k v(l)\mbox{, for all }1\le k \le L-1.
\end{equation}
The algorithm consists of generating each of the distributions
(\ref{e:genclass}--$c$) separately:
\begin{enumerate}
\item Compute the classical integral (\ref{e:genclass}).   This is
needed for the normalization of the next Metropolis step and,  by
(\ref{e:classbound}), it can be used also for determining if a
computation of the full term is necessary.
\item Use e.g.\ a Metropolis algorithm to generate the classical
distribution in (\ref{e:genclass}) for $y$.
\item \label{it:genRW} Generate the distribution for $v$ by applying
a Metropolis check with normal distributed trials  for each coordinate
of $v$.  This can be done independently of the generation of $y$ and,
in fact,  each classical coordinate can also be generated separately.
This problem is equivalent to the generation of  a self-recurring
one-dimensional random walk of $L$ steps, the cumulative sums of $v$
corresponding to the positions of the walker.  Note that the constants
in (\ref{e:genqf}) have already been chosen so that the result is a
probability distribution.
\item Apply a Metropolis check implied by (\ref{e:genrest}) for the
generated values of $y$ and $v$.
\end{enumerate}

In practice, it is best to store the values of $x(k)$ and to use a
separate routine for the generation of  the trials for quantum
fluctuations of each particle.  The idea behind the algorithm is to
generate a path of possible quantum fluctuations for a classically
distributed particle and to test for the acceptance of the whole
fluctuation path simultaneously.   Since $P_L$ does not enter this
acceptance test at all, this avoids the large rejectance rate from
which a local change made in just one copy suffers.  But as the
proposed change is not local, the evaluation of the  last Metropolis
check is made more difficult and  time-consuming---this should be more
than compensated by the improved acceptance rate.

Let us also briefly comment on the generation of the classical
distribution when $s$ is not the identity permutation.  Then $s$ has
at least one cycle of length $\ell>1$ and assume, for simplicity, that
it is composed of  the first particle labels.  By the same change of
variables as before, i.e.\ by defining 
$\bi{y}_i=\sum_{l=1}^i\bi{u}_l$, we get then
\[
\sum_{i=1}^\ell |\bi{y}_{s(i)}-\bi{y}_i|^2 = \sum_{i=2}^\ell
 |\bi{u}_i|^2 + \bigg|\sum_{i=2}^\ell \bi{u}_i\bigg|^2.
\]
This is independent of $\bi{u}_1$, i.e.\ of one of the particle
positions.  On the other hand, it also implies that the distance
between  any two particles is of the order of $\sqrt{\beta'}$.  It is
thus possible to think of each $\ell$-cycle as describing a cluster of
$\ell$ particles which move together in the external potential.
Similarly, especially for higher temperatures, it is best to generate
the combinations $\bi{u}$ from the Gaussian distribution and use the
potential only for $\bi{u}_1$ and, of course, for the Metropolis
check.

\section{Lattice evaluation of Gaussian traces}\label{sec:gauss}

The Gaussian ensemble has been introduced in \cite{jml:gauss}, where
its connections to the canonical and the microcanonical ensemble were
also illustrated.  The ensemble is defined by the unnormalized density
operator
\[
\tihep(E) = \frac{1}{\sqrt{2\pi \vep^2}}
  \exp\!\bigg[-\frac{1}{2\vep^2} (\ham-E)^2\bigg].
\]
It was proven there that for any observable $\A$ which satisfies the
condition $\Tr(|\ope{A}| e^{-\beta\ham})<\infty$ for all $\beta>0$, we
have the following integral representation
\begin{equation}\label{e:inta}
\Tr\!\left(\ope{A} \tihep(E)\right) =
  \int_{\beta-\rmi\infty}^{\beta+\rmi\infty} {\rmd z\over 2\pi\rmi}\,
  \rme^{\half\vep^2 z^2 + z E} \Tr\!\left(\ope{A}\rme^{-z\ham}\right),
\end{equation}
where $\beta$ is a positive constant.

For numerical computations, the proper choice of $\beta$ is very
important.  Suppose we wish to examine the behaviour of the system
near the energy scale $E_0$ using the energy resolution  $\vep< E_0$.
At $E=E_0$, it was shown in \cite{jml:gauss} that the best choice for
$\beta$ is the unique positive solution to the equation
\begin{equation}\label{e:defE0}
 E_0 + \beta \vep^2 = \barh \equiv
 \frac{\Tr(\opm\ham \rme^{-\beta\ham})}{\Tr(\opm\rme^{-\beta\ham})}.
\end{equation}
In practice, it is usually easier to begin with  the temperature
$1/\beta$ and then solve $E_0$ from equation (\ref{e:defE0}).  The
natural energy resolution parameter  is then $\bep=\beta\vep$ and as
the energy parameter, it will be easiest to use the scaled difference
$\be=\beta(E-E_0)$.  Similarly, the moments are most naturally
computed as
\[
\beta^k \gmean{(\ham-E-\beta\vep^2)^k}_{E,\vep} =
 \gmean{(\beta(\ham-\barh)-\be)^k}_{E,\vep}.
\]
We shall derive the Gaussian lattice expressions  for this set of
moments---all others can, of course, be computed from these.

With the present assumptions, $\A=\opm$ satisfies the condition for
the use of the integral representation (\ref{e:inta}).  Therefore, for
any real $t$ and $\beta>0$ the following holds,
\begin{equation}\label{e:intpt}
\fl \Tr\!\left(\opm \rme^{t(\ham- E-\beta\vep^2)} \tihep(E)\right) =
  \int_{\beta-\rmi\infty}^{\beta+\rmi\infty} {\rmd z\over 2\pi\rmi} 
  \,\rme^{\half\vep^2 t^2+t\vep^2 (z-\beta)} 
  \rme^{\half\vep^2 z^2 + z E} \Tr\!\left(\opm\rme^{-z\ham}\right).
\end{equation}
Applying proposition \ref{th:anint} then shows that both sides can be
analytically continued to entire functions and that for all $k\ge 0$,
\begin{eqnarray*}
\fl \Tr\!\left(\opm(\ham-E-\beta\vep^2)^k \tihep(E)\right) =
  \frac{\rmd^k}{\rmd t^k} \left.\Tr\!\left(\opm 
  \rme^{t(\ham-E-\beta\vep^2)} \tihep(E)\right)\right|_{t=0} \\ 
\lo=\int_{\beta-\rmi\infty}^{\beta+\rmi\infty} 
  {\rmd z\over 2\pi\rmi}\,\frac{\rmd^k}{\rmd t^k}
  \left.\rme^{\half\vep^2 t^2+t\vep^2 (z-\beta)}\right|_{t=0} 
  \rme^{\half\vep^2 z^2 + z E} \Tr\!\left(\opm\rme^{-z\ham}\right).
\end{eqnarray*}
The derivative on the right hand side can be computed from the
generating function of the Hermite polynomials $H_k$,
\begin{eqnarray*}
\fl {\rmd^k\over \rmd t^k} 
  \left.\rme^{\half \vep^2 t^2 + t \vep^2 (z-\beta)}\right|_{t=0} 
  = \left( \rmi \vep/\sqrt{2} \right)^k
  H_k\!\left(-\rmi (z-\beta) \vep/\sqrt{2}\right) \\ 
\lo= \vep^k \sum_{j=0}^{\left\lfloor k/2\right\rfloor} 
  { k!\over 2^j j! (k-2 j)! } [(z-\beta) \vep]^{k-2 j}
\end{eqnarray*}

Parameterizing the integral as $z=\beta(1+\rmi \alpha)$ and using the
scaled variables $\bep$ and $\be$ we have the result
\begin{eqnarray*}
\fl \beta^k\Tr\!\left(\opm(\ham-E-\beta\vep^2)^k \tihep(E)\right) \\
\lo= \beta\bep^k \rme^{-\half \bep^2+ \be +\beta\barh}
  \sum_{j=0}^{\left\lfloor k/2\right\rfloor}  
  { k!\over 2^j j! (k-2 j)! } I_{k-2 j}(\be,\bep;\beta),
\end{eqnarray*}
where
\begin{eqnarray*}
I_k(E,\vep;\beta) =  \rmi^k \int_{-\infty}^{\infty}\!
 \frac{\rmd\alpha}{2\pi} (\vep\alpha)^k  
 \rme^{-\half\vep^2\alpha^2+\rmi\alpha (E+\beta\barh)}
 \Tr\!\left(\opm\rme^{-\beta(1+\rmi\alpha)\ham}\right).
\end{eqnarray*}
Let us next define the function
\[
g(\alpha;\beta) = \rme^{\rmi\alpha\beta\barh}
  \Tr\!\left(\opm\rme^{-\beta(1+\rmi\alpha)\ham}\right),
\]
using which we have for even $k$,
\begin{equation}\label{e:defeven}
I_k(E,\vep;\beta) = (-1)^{k/2}\, \re\!\! \int_{-\infty}^{\infty}\!
 \frac{\rmd\alpha}{2\pi} \rme^{\rmi\alpha E} (\vep\alpha)^k
 \rme^{-\half \vep^2\alpha^2} g(\alpha;\beta),
\end{equation}
and for odd $k$,
\begin{equation}\label{e:defodd}
I_k(E,\vep;\beta) =  (-1)^{(k+1)/2}\, \im\!\!
 \int_{-\infty}^{\infty}\!  \frac{\rmd\alpha}{2\pi} 
 \rme^{\rmi\alpha E} (\vep\alpha)^k  
 \rme^{-\half \vep^2\alpha^2} g(\alpha;\beta).
\end{equation}
This shows that if we know the function $g$ in some suitably dense set
of points, we can approximate $I_k$ for any $k$, $\bep$ and $\be$ by
multiplying the known values by the weight function and then
performing a discrete Fourier-transform.

Thus $g(\alpha;\beta)$ contains all information even about the
arbitrarily high energy end of the spectrum and it should not come as
a surprise that its computation  is not easy in general.  Fortunately,
it has a lattice approximation and there is a relatively
straightforward Monte Carlo algorithm  for the evaluation of the
lattice integral.

By theorem \ref{th:latconv}, the lattice approximations
\begin{equation}\label{e:defgL}
g_L(\alpha;\beta) =  \rme^{\rmi\alpha\beta\barh}  
  \Tr\!\left(\opm \ope{T}(\beta(1+\rmi\alpha)/L)^L\right)
\end{equation}
converge to $g(\alpha;\beta)$ as $L\to\infty$ and they are all
uniformly bounded in $\alpha$.  If we define  $I_k(E,\vep;\beta,L)$ by
replacing $g$ by $g_L$ in (\ref{e:defeven}) and  (\ref{e:defodd}), we
can thus rely on the dominated convergence theorem and conclude that
\[
\beta^k \gmean{(\ham-E-\beta\vep^2)^k}_{E,\vep} = \lim_{L\to\infty}
\bep^k \sum_{j=0}^{\left\lfloor k/2\right\rfloor}  
 { k!\over 2^j j! (k-2 j)! }  
 \frac{I_{k-2 j}(\be,\bep;\beta,L)}{I_0(\be,\bep;\beta,L)},
\]
where $\be = \beta (E+\beta \vep^2-\barh)$ and $\bep=\beta\vep$.
Since the Gaussian partition function measures directly the density of
states, it is of a special interest and, by the above results, it has
the lattice approximation
\[
\Tr\!\left(\opm \tihep(E)\right) =  \lim_{L\to\infty} 
 \rme^{\half\beta^2 \vep^2 + \beta E} \beta I_0(\be,\bep;\beta,L).
\]

\subsection{Monte Carlo algorithm for the Gaussian integrals}

Let us now go through the lattice algorithm for the evaluation of
$g_L$ when $\alpha$ and $\beta>0$ are given.  The definition
(\ref{e:defgL}) translates into the lattice integral
\begin{eqnarray}
\fl g_L(\alpha;\beta) = \frac{1}{N!} \sum_{s\in S_N} (\pm )^s
  \rme^{\rmi\alpha\beta\barh} \int\!\rmd^{nL} x\,
  K_L(sx(L),x;\beta(1+\rmi\alpha)) \nonumber\\ 
\lo= \frac{1}{N!} \sum_{s\in S_N} (\pm )^s 
  (1+\alpha^2)^{\frac{Ln}{4}} \int\!\rmd^{nL} x\,  
  \biggl( {L\over 2 \pi \beta'} \biggr)^{{L n\over 2}} 
  \exp\!\bigg[-\frac{1}{\beta'} P_L - \beta V_L\bigg] \nonumber\\ 
\times \exp\!\bigg[\rmi\alpha\bigg( \beta\barh-\beta V_L +
  \frac{1}{\beta'} P_L - \case12 Ln 
  \frac{\arctan \alpha}{\alpha}\bigg)\bigg],\label{e:glat}
\end{eqnarray}
where $\beta'$ is a short-hand for the combination  
$\beta (1+\alpha^2)$ and $P_L=P_L(sx(L),x)$ 
and $V_L=V_L(x)$ as before.  In
addition, since the kernel of $\T$ is obtained by analytic
continuation, the arcus tangent in the above should be chosen from the
primary branch, i.e.\ $\arctan \alpha \in (-\case12\pi,\case12\pi)$.

Now we can also finally explain why we wanted to include the term
$\beta\barh$ into the lattice integration. Suppose, for simplicity,
that of the possible classes of permutations one dominates over the
others---this  is natural if the quantum behaviour of the system comes
from the formation of correlated particle clusters of fixed size for
the temperature in question (we saw in section \ref{sec:canalg} that
when a permutation is decomposed into cycles, each cycle can naturally
be identified as such a correlated cluster of particles).  Then for
this permutation and for small $\alpha$, the last term in parentheses
in  (\ref{e:glat}) has an almost zero expectation value under the
distribution in front of it---this follows from a comparison with the
definition of $Q_1$ in (\ref{e:defQ1}).  Since the
$\alpha$-integration is concentrated near zero, choosing $\beta\barh$
as the energy origin thus gives the smallest oscillations for the
integrand in (\ref{e:glat}).  This also shows, that for a fixed
lattice size $L$, it is best to use the value evaluated from the same
lattice for $\barh$, a continuum extrapolation is not necessary or
desirable.

The first term in the lattice integral in (\ref{e:glat}) gives simply
a canonical distribution which can be generated by the algorithm
explained in section \ref{sec:canalg}.  The evaluation of the
remaining exponential is then a simple computation of trigonometric
functions.

Computation of $g$ for all $\alpha$ would then enable the solution of
the full microcanonical spectrum of the Hamiltonian.  As the energy is
increased, the dependence of the position of the spectral lines
depends on ever finer structure of the potential. Since, eventually, a
computer cannot hold the values of the potential at the accuracy
needed, there must be a catch somewhere.  In our case, this is most
easily seen in the prefactor $(1+\alpha^2)^{\frac{Ln}{4}}$ which must
the cancelled by the result from the lattice integral---otherwise $g$
would diverge in the continuum limit.  This means that as $\alpha$ is
increased, the oscillations of the lattice observable increase rapidly
and the maximum value of $\alpha$ that can be computed this way is
limited by the number of Monte Carlo sweeps feasibly performed and by
the resulting inaccuracy of the result from the lattice integration.

\section{An example: charged particles in a quadratic potential} 
\label{sec:coulomb}

We now want to check how well the algorithm works in practice.  We
shall consider particles living in a three-dimensional world and
bound by a harmonic potential.  After we have checked that the
algorithm works for this analytically solvable case, we shall add a
Coulomb interaction between the particles.  In other words, we shall
consider the potential
\[
V(x) = \sum_{i=1}^N \half\omega^2 \bi{x}_i^2 + \sum_{i=1}^N
 \sum_{j=1,j\ne i}^N \half  \frac{q^2}{|\bi{x}_i-\bi{x}_j|},
\]
where $N$ gives the number of particles, $\omega$ is the binding
frequency and $q$ denotes the charge of one particle.

The scaling properties of the potential part of the action are now
simple,
\[
\beta V_L(\sqrt{\beta} x) = \beta^2 V_Q(x) + \sqrt{\beta} V_C(x),
\]
where
\[\fl
 V_Q(x) = {1\over L} \sum_{k=1}^L \sum_{i=1}^N \half\omega^2
  \bi{x}_i^2(k) \quad\mbox{and}\quad V_C(x) = 
  {1\over L} \sum_{k=1}^L \sum_{i=1}^N \sum_{j=1,j\ne i}^N \half
  \frac{q^2}{|\bi{x}_i(k)-\bi{x}_j(k)|}.
\]
By first scaling the temperature $\beta$ away from the kinetic term,
and by subsequent differentiation we can thus derive a second set of
observables for the computation of the energy moments.  The energy
expectation value can now be measured either by using $Q_1$ in
(\ref{e:defQ1}) or by using the observable
\begin{equation}\label{e:defQbark}
\bar{Q}_1 = 2 V_Q + \half V_C,
\end{equation}
which has been rescaled back to the original lattice variables.
Similarly, the second moment is given by (\ref{e:defQ2}) or by
\[
\bar{Q}_2 = \bar{Q}_1^2 + \frac{1}{\beta} 
  ( 2 V_Q -  \frac{1}{4} V_C )
\]
and the other observables $\bar{Q}_k$ could be computed equally
easily.  Note that, again, simply taking the second power  of the
observable giving the energy expectation value does not yield the
correct result.

The advantage of using both of these two kinds of observables comes
from their very different dependence on the lattice kinetic energy.
For large $L$ the second set of observables is better, since it does
not contain any cancellation of two large numbers as is necessary for
the first set of observables.   For all values of $L$, both methods
should nevertheless yield mutually consistent results and we, in fact,
used this for fine-tuning  and checking of the Monte Carlo algorithm.

\begin{table}
\caption{First six moments $\mean{\beta^k (\ham-\mean{\ham})^k}$  of a
two-particle system in a three-dimensional space. Only the 
$s={\rm id}$ term has been considered, 
i.e.\ the particles have been assumed
to be distinguishable, and the binding frequency is $\omega=1$.  The
results were obtained from an eight-step lattice  using either $Q_k$
as defined in equation (\protect\ref{e:defQk}) or $\bar{Q}_k$ as
defined in section \ref{sec:coulomb}.   For the non-charged case, the
known continuum value is also given. \label{t:canmom} }
\begin{indented}
\lineup
\item[]\begin{tabular}{@{}*{10}{l}} \br
 $\beta$ & $q$ & obs. & $\mean{\beta\ham}$ & $k=2$ &  $k=3$ & $k=4$ 
  & $k=5$ & \0$k=6$ \\ \mr  
 1.0 & 0 & $Q$ & 6.483(3) & 5.59(2) & 11.7(2) & 131(2) 
  & 778(25) & \07100(290) \\ 
 & & $\bar{Q}$ & 6.481(4) & 5.49(3) & 11.4(3) & 123(5) 
  & 790(100) & \08800(2500) \\ 
 & & c.l.{} & 6.4919 & 5.524 & 11.95 & 127.6 & 804.4 & \07664 \\ 
 & 2 & $Q$ & 8.735(3) & 5.03(2) & 10.8(2) & 110(2) 
  & 647(23) & \06150(430) \\ 
 &  & $\bar{Q}$ & 8.665(4) & 4.69(3) & \09.8(3)& \098(7) 
  & 730(170) & \09400(4000) \\    
 0.1 & 0 & $Q$ &  6.000(3) & 5.99(2) & 11.8(2) & 147(2) 
   & 832(27) & \08930(440) \\ 
 & & $\bar{Q}$ & 6.000(2) & 5.970(12) &11.87(15) & 143(2) 
   & 886(40) & \09340(790) \\ 
 & & c.l.{} & 6.0050 & 5.995 & 12.00 & 143.8 & 863.4 & \08629 \\ 
 & 5 & $Q$ & 6.583(3) & 5.82(2) & 11.6(2) & 139(2) 
   & 852(26) & \08690(370) \\ 
 &  & $\bar{Q}$ & 6.585(2) & 5.79(2) & 11.8(2) & 138(3) & 907(65) 
   & 10300(1500) \\ \br
\end{tabular}
\end{indented}
\end{table}

We have presented the results of the Monte Carlo computation of the
energy moments of a system of two distinguishable  three-dimensional
particles at two different temperatures in Table \ref{t:canmom}.   The
results were computed performing $3\cdot 10^6$ sweeps on an
AlphaStation 1000 computer which took a few hours of computer time.
The error estimates for the moments were computed using the
bootstrap-method and the number of sweeps was enough for obtaining a
clean signal for all of the measured six moments.

In both cases, we used a lattice with $L=8$ steps and it is clear that
the results are very close to the continuum limit.  In fact, the main
difference to the classical $L=1$ results comes already at the
addition of one classical copy for the quantum fluctuations, i.e.\
already at $L=2$.   As expected,  the convergence to the continuum
limit is also faster for smaller values of $\beta$.

We have included the $\beta=1$ case as a warning about the possible
systematical errors induced by the Monte Carlo algorithm.   It is
clear that the two lattice observables at $\beta=1$ do not yield as
mutually consistent results as at $\beta=0.1$.  This is most clearly
seen from the energy expectation value of the charged case when the
two results are not consistent within their statistical errors when
$\beta=1$, but they {\em are}\/ consistent for $\beta=0.1$  even
though we have increased the charge for the second case.

The source of these systematical errors is in the \apriori\
distribution which is used for getting samples for the Metropolis check.
If the \apriori\  distribution does not concentrate to the correct
region of the configuration space, the resulting probability
distribution  in this region will be coarser than suggested by the
number of sweeps used and the results, likewise, less accurate.

For instance, the differences in the non-charged case are caused by
the use of quantum fluctuations as the \apriori\  distribution
although we are already in the low-temperature region $\beta\to\infty$
where the lowest energy state dominates.  This could be remedied by
using also sweeps with the usual Monte Carlo sampling where the
binding potential $\e^{-\beta V_L}$ serves as the \apriori\
distribution.  The discrepancy in the low-temperature charged case is
more problematic, since it is caused also by the singularity of the
Coulomb distribution.  The cure in the second case would be replacing
the Coulomb distribution with some smooth approximation of it, but the
effect of this change would then have to be analyzed separately.   Of
course, it is always possible simply to increase the number of sweeps
until the final probability distribution is smooth enough, but this
might not always be feasible.

\begin{figure}
\begin{center}
\mbox{\epsfysize=9.5cm\epsfbox{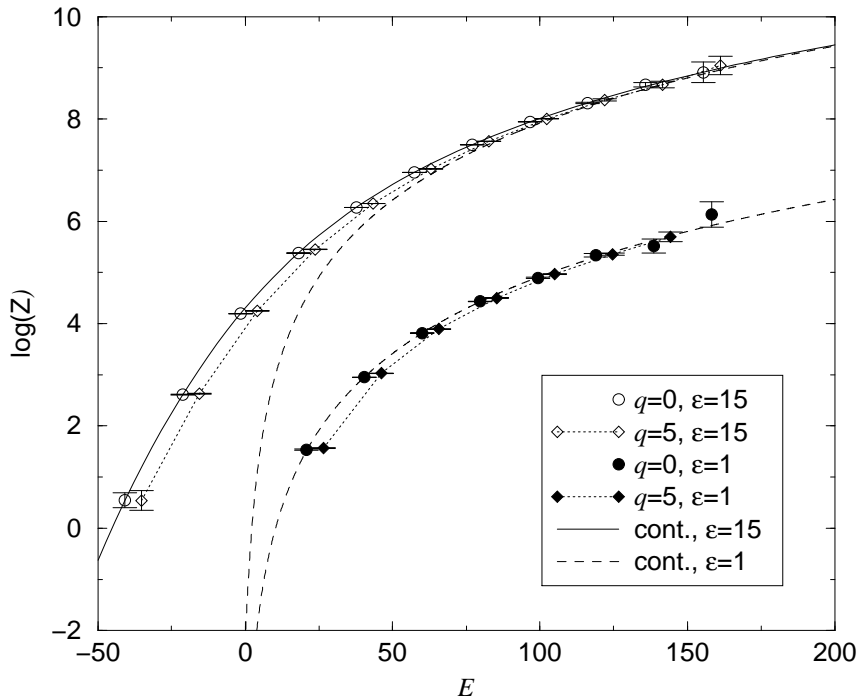}}
\end{center}
\vspace*{-.5cm}
\caption{The base ten logarithm of the Gaussian partition function
obtained using a four-step lattice with $\omega=1$ and $\beta=0.1$ and
the two shown values of the energy resolution.  For the sake of
clarity, the results for $\vep=1$ are plotted three orders of
magnitude smaller than they were measured.  The solid and dashed lines
represent the continuum results for $q=0$; the dashed line has been
plotted twice, both at its correct and  reduced
location.\label{f:gauss}}
\end{figure}

We also measured the Gaussian partition function, i.e.\ the density of
states, for this system.  The measurements were performed by computing
$g_L(\alpha;\beta)$ for $|\alpha|\le 1.6$ at intervals of $0.05$ on a
four-step lattice using $\beta=0.1$.  Samples were then taken from the
measured values with the assumption of normally distributed errors and
a discrete Fourier-transformation with the correct weight was
performed to each of these samples.  We have given the  stable results
of this analysis in figure \ref{f:gauss} for two different values for
the energy resolution, $\bep=1.5$ and $\bep=0.1$.  It is interesting
that we were able to get accurate results also for the second case,
for which the effect of the Gaussian weight is almost negligible.

Otherwise, the results behave exactly as expected.  The most accurate
results are obtained near $\be=0$, which corresponds to $E\approx 40$
in the first case and to $E\approx 60$ in the second case.  The
accuracy detoriates as the energy difference increases, faster in the
case of smaller energy resolution.  The addition of the Coulomb term
also seems to affect the density of states only at the low energy
spectrum.

The non-charged results were also compared to  their continuum
counterparts which were computed using the known partition function
\[
\Tr\!\left(\e^{-z\ham}\right)_{q=0} = 
  \frac{1}{\left(2\sinh(z\omega/2)\right)^{N d}}.
\]
If we remember that we used only an $L=4$ lattice, the results
presented in figure \ref{f:gauss} are surprisingly accurate.  When we
later checked the difference between the classical $L=1$ results and
the full quantum results at $\beta=0.1$ for the values of $\alpha$ we
used, the differences were almost negligible.  This seems to imply
that, at least for the harmonic potential, the classical density of
states does give a highly accurate measurement of the high energy
density of states  of the quantum system.

In contrast, we were not able to measure the discrete structure of the
energy spectrum by measuring $g_L$ since this would have needed
impractically high values for $\alpha$.  We also checked that for
detecting the discrete spectrum, the use of large values of $L$ is
essential.  Thus the present method is suitable only for measurement
of quantities which do not depend directly on the  discrete energy
spectrum, but only on the density of states.

\section{Prologue: improvements and a brief comparison with
existing methods}\label{sec:prologue}

A number of references came to our attention after the Monte Carlo
simulations were finished and in this section we shall suggest some
improvements motivated by them.  We first present a better algorithm
for the generation of the self-recurrent random walk inspired by the
work done mostly by Doll and Freeman, see e.g.\ \cite{dollf84b}.   The
second subsection contains a discussion and  references to other
possible improvements.

\subsection{Generation of the random walk by Fourier components}

Suppose we make, instead of (\ref{e:deful}), a change of variables
defined by a discrete Fourier sine-transform,
\[
a(k) = \frac{2}{L} \sum_{j=1}^L [x(j)-c(j)] 
 \sin\!\left(\pi\case{k}{L} j\right)
 \mbox{, for all }1\le k \le L-1.
\]
The following ``orthogonality'' relations, which can be found from
\cite{hansen} or computed directly by expressing sines as
exponentials, are valid for the discrete sine-transform,
\begin{eqnarray*}
\sum_{j=1}^L \sin\!\left(\pi \case{k}{L} j\right) 
 \sin\!\left(\pi\case{k'}{L}j\right) =  
 \frac{L}{2} \delta_{kk'}, \\ 
 \fl \sum_{j=1}^L \sin\!\left(\pi \case{k}{L} j\right) 
 \sin\!\left(\pi\case{k'}{L} (j-1)\right) =  
  \cases{ \frac{L}{2}\cos\!\left(\pi\case{k}{L}\right) & 
     for $k=k'$ \\ \frac{1-(-1)^{k+k'}}{2}
        \frac{\sin\!\left(\pi \case{k}{L}\right) 
  	 \sin\!\left(\pi\case{k'}{L}\right)}{
	    \cos\!\left(\pi \case{k}{L}\right)-
	    \cos\!\left(\pi \case{k'}{L}\right) } 
   & for $k\ne k'$ \\ }, \\
\end{eqnarray*}
where we have assumed that both $k$ and $k'$ belong to
$\set{1,2,\ldots,L-1}$.  Using these formulae it  is easy to prove
that the linear transformation which defines the change of variables
is always invertible with an inverse
\begin{equation}\label{e:deffxj}
x(j) = c(j) + \sum_{k=1}^{L-1} a(k) 
 \sin\!\left(\pi\case{j}{L} k\right) \mbox{, for all }0\le k \le L,
\end{equation}
and that with the choice $c(k)= (1-\case{k}{L}) sy + \case{k}{L} y$
the kinetic term becomes simply
\begin{eqnarray*}
P_L = {1\over 2} |y-sy|^2 + \sum_{k=1}^{L-1} |a(k)|^2 L^2
  \sin^2\!\!\left(\case{\pi k}{2 L}\right).
\end{eqnarray*}
Since equation (\ref{e:genqf}) integrates to unity, we can then also
deduce the Jacobian of this change of variables and conclude that
(\ref{e:genqf}) can be replaced by
\begin{equation}\label{e:genfour}
%\fl
 \int\!\!\frac{\rmd^{n(L-1)}a }{ (2 \pi)^{n (L-1)/ 2} }
  \prod_{k=1}^{L-1} \frac{1}{\sigma_k^n} \exp\!\biggl[-\frac{1}{2}
  \sum_{k=1}^{L-1} |a(k)|^2/\sigma_k^2\biggr],
\end{equation}
where
\begin{equation}\label{e:defsigmak}
 \sigma_k^2 = \frac{\beta'}{2  L^2 
  \sin^2\!\!\left(\case{\pi k}{2 L}\right)}.
\end{equation}

Therefore, we can replace the Monte Carlo step suggested in item
(\ref{it:genRW}) in section \ref{sec:canalg} by a generation of the
path $x(k)$ via the coefficients $a(k)$ in equation (\ref{e:deffxj}).
Since all $a(k)$ have a normal distribution, their generation is fast
and straightforward.   Note also that the discrete Fourier
transformation need not slow down the algorithm, since all necessary
trigonometric functions  can be computed before entering the Monte
Carlo loop.

\subsection{Discussion about the Fourier method and further improvements}

Using Fourier-transformation of the path to express quantum
statistical path-integrals was suggested already by Feynman
\cite{FHibbs}.  Later, Doll and Freeman \cite{dollf84b} used  a form
with a cut-off for the number of Fourier modes in a Monte Carlo
simulation of energies of quantum mechanical systems and they found a
performance better  than in a conventional Metropolis Monte Carlo.
However,  the convergence of energy expectation values  was found to
be non-monotonic for this ``primitive Fourier method'' and a
relatively high number of Fourier modes, $k_{\rm max} \approx 50$,
was necessary before the continuum value was approached
\cite{dollf99,dfcomment99}.

In our simulations, we did not see any non-monotonicity of the
convergence of the lattice energy expectation values and,  in
addition, we found that a relatively coarse lattice  was sufficient
for the computation of nearly continuum results.  Emboldened by these
results we now suggest that using  the version of the Fourier
transformation given in the previous section should solve the problems
related to the approach to the continuum limit that are manifest in
the primitive Fourier method.

In other words, if the ``Matsubara frequencies'' used in the primitive
Fourier method are replaced by those derived in the previous section,
i.e.\ if we replace
\[
 \sigma_k^2 = \frac{2 \beta'}{(k \pi)^2} \qquad\mbox{ by }\qquad
 \sigma_k^2 =  \frac{\beta'}{2 L^2 
  \sin^2\!\!\left(\case{\pi k}{2 L}\right)},
\]
then the Fourier-method should converge for smaller values of 
$k_{\rm max} = L-1$.  
Similarly, the integrals needed in the previous energy
estimators should probably be replaced by their discretized lattice
versions.  This idea is not new \cite{runge88},  but it has not been
widely used, either.

Since we did not use as complicated a system as in the references, it
would now be necessary and interesting to repeat the test of the
convergence and the comparison of computing times for the different
methods that were done in \cite{dollf99} and \cite{chakra98},
respectively.  For this comparison, we would like to point out that
the differentiation which we used for producing the second set of
observables, $\bar{Q}_k$ in section \ref{sec:coulomb}, would for a
generic potential lead to what is  essentially the virial estimator
defined in \cite{dollf84b}.  Note, however, that our method also
yields similar estimators for all other energy moments, not just for
the first one.  In addition, both of our lattice observables always
give the same result on the same lattice, whereas the previous ones
agree only in the continuum limit.

We can now also explain why the Fourier method is better than the
conventional lattice Monte Carlo: using the Fourier coefficients as
the \apriori\ distribution gets rid of the kinetic term which we
suggested was the culprit for the slow convergence of the conventional
Monte Carlo algorithm for large $L$.  We can also explain why
typically no estimators are given for energy moments beyond the first
one: since each power of the kinetic term requires a separate
renormalization, guessing the correct lattice observables from
continuum expressions is difficult and the effect of using wrong
constants can easily be drastic.  In addition, observables for the
higher moments cannot be reduced to ``coordinate-space observables''
(i.e.\ observables living in only one time-slice) as is the case for
the virial estimator of the energy for distinguishable particles.

Since the changes we suggest here make the Fourier lattice integrals
{\em equivalent}\/ to the discrete time-step integrals, the methods
used for improving the performance of the earlier methods should apply
equally easily here.  One could e.g.\ employ classical cluster
algorithms \cite{chand80} or use  smeared or effective potentials
\cite{dollm79}.   One interesting application which we did
not discuss  here at all is in the computation of real-time
correlation functions, for this see e.g.\ \cite{thiru83,dollf89a}.

Although the algorithm for the computation of Gaussian traces is not
practical for resolving the discreteness of the energy spectrum, the
lattice formulae derived  are valid for arbitrarily fine energy
precision.  By following the steps which where used in
\cite{dollf89b}, we could derive a lattice formulation with a
complicated oscillating kernel which could be used for solving also
the low-energy discrete spectrum.   How far in energy and for how
complicated systems this kernel can be used in practise needs testing.

\section{Conclusions}

We have presented a lattice Monte Carlo algorithm for the computation
of energy moments in the canonical Boltzmann-Gibbs ensemble  for a
system of non-relativistic quantum mechanical particles.  In fact,
none of our results would have changed if we were to add a bounded
classical observable $A(x)$ to the trace.  Thus the present method can
be used also for the evaluation of microcanonical corrections   to the
canonical expectation values of such observables as is explained in
\cite{jml:gauss}.   The overall performance of the algorithm,
especially for high temperatures, was very good.  In addition to the
algorithm, we presented rigorous limits which can be used for a
quantitative estimation of when and how the indistinguishability of
particles will begin to affect the results.

We also proposed a lattice formulation  which can be used for the
evaluation of Gaussian expectation values and the Gaussian partition
function of these systems.  The algorithm we gave for the computation
of the lattice integrals was shown to yield correct information about
the density of states even for quite coarse lattices.  If knowledge
about the discrete energy spectrum is required, then using analytical
methods or a modification of the algorithm seems to be necessary.  The
present method has the advantage that other non-canonical ensembles in
addition to the Gaussian one, as well as many different energy
resolutions, can be analyzed from the same lattice data.

In both cases, we saw that the classical ensembles work surprisingly
well and we noted how the intuitive understanding of the classical
case can help in the computation of  the full quantum results.  The
main use of the Gaussian lattice methods would likely to be in the
detection of phase transitions.  Although it is necessary to increase
the number of particles to infinity before a clean signal for a phase
transition appears, such effects should be visible for a finite number
of particles as well.

\ack I am indebted for the motivation and advice provided by
M.~Chaichian.  I also wish to thank A.~Kupiainen and C.~Montonen for
their comments on the manuscript.

\appendix

\section{Analytic integrals}\label{sec:anint}

The following result is a straightforward consequence of textbook
results, but  since it is central to all the results presented in the
text, we give a brief derivation for it. The main content of the
theorem is that an analytically parameterized integral is an analytic
function of the parameter if the integrand is uniformly dominated by
an integrable function in every compact subset of the parameter space
and that in this case it is also always possible to take
differentiation of the parameter inside the integral.

\begin{proposition}\label{th:anint}
Let $\Omega$ be an open subset of the complex plane and let $X$  be a
measure space which is $\sigma$-finite with respect to a positive
measure $\mu$.  Consider a function $F: X\times\Omega \to \C$, which
is measurable  in the product space and which defines an analytic
function $z\mapsto F(x,z)$ for almost every (w.r.t.\ $\mu$) $x$.  If
for every compact subset $K$ of $\Omega$ there is a function 
$g_K\in L^1(\mu)$ such that 
$|F(x,z)|\le g_K(x)$ for almost every $x$ and for
every $z\in K$, then the function $f: z\mapsto \int d\mu(x) F(x,z)$ is
well-defined and analytic in $\Omega$.  Then also for all $z\in\Omega$
and positive integers $k$,  
${d^k\over dz^k} f(z) = \int d\mu(x) {d^k\over dz^k}F(x,z)$.
\end{proposition}

\begin{proof} 
$f$ is clearly
well-defined for all $z\in\Omega$.  By the dominated convergence
theorem it is also continuous and then an application of the Morera's
theorem shows that it is analytic.  Using Cauchy's integral formula
for the derivatives of $f$, followed by an application of Fubini's
theorem then leads to the given expression for the derivatives. For
more details see \cite{rudin:rca}.
\end{proof}

When using the proposition here, the $\sigma$-finiteness is obvious
since we use it only for the counting measure on the positive integers
and for Lebesque measures on Euclidean spaces.  The measurability of
$F$ in the product space is also trivial since  the functions we are
dealing with will be continuous.

\section{Proof of the lattice convergence theorem}\label{sec:latproof}

We prove here how theorem \ref{th:latconv} follows from the
assumptions (\ref{it:Vcon}) and (\ref{it:potcon}) given in section
\ref{sec:notations}.  The proof consist of four parts: we first show
that the $k=0$ trace is an analytic function which works as a
generating function for the other powers.  We then derive an integral
representation for the lattice traces and use this to prove their
analyticity.  Next we show that the lattice traces are uniformly
bounded on compact subsets with a bound given in terms of the
classical partition function.  Since the traces converge for positive
values of the parameter, theorem \ref{th:latconv}  follows from an
application of  the Vitali convergence theorem \cite{vitalict2} and
properties of analytic functions.

However, let us begin by defining the following two  functions of a
lattice configuration $x=(x(1),\ldots,x(L))$,
\begin{eqnarray*}
P_L(b,x) = {L\over 2}\sum_{k=1}^L |x(k)-x(k-1)|^2 
  \mbox{, with }x(0)=b\mbox{, and} \\ 
V_L(x) = {1\over L}\sum_{k=1}^L V(x(k)).
\end{eqnarray*}
In the proof and in practical computations the following bound for
their exponentials will become useful.
\begin{lemma}\label{th:ikbound}
For any permutation $s\in S_N$ and for all $L\ge 1$ and $\beta>0$,
\begin{eqnarray}\label{e:lemeq}
\fl 
\int\!\rmd^{nL} x\,
 \biggl( {L\over 2 \pi \beta} \biggr)^{{L n\over 2}} 
 \exp\!\bigg[ -\frac{1}{\beta} P_L(sx(L),x) - 
  \beta V_L(x)\bigg]  \nonumber\\ 
\lo\le \int\!\! \frac{\rmd^n y}{(2\pi\beta)^{n/2}}\,  
 \exp\!\bigg[{-\frac{1}{2\beta}
   |sy-y|^2-\beta V(y)}\bigg] \le \Zcl(\beta).
\end{eqnarray}
\end{lemma}
\begin{proof}
By the relation between arithmetic and geometric means, we always have
$\exp[-\case1L \sum_{k=1}^L \beta V(x(k))] \le  \case1L \sum_{k=1}^L
  \exp[-\beta V(x(k))]$.  Applying this and the rules of Gaussian
integrals then yields
\begin{eqnarray*}
\fl 
\int\!\rmd^{nL} x\, 
 \biggl( {L\over 2 \pi \beta} \biggr)^{{L n\over 2}} 
 \exp\!\bigg[ -\frac{1}{\beta} P_L(sx(L),x) - \beta
   V_L(x)\bigg] \\ 
\lo\le  \frac{1}{L} \int\!\rmd^n y \, \rme^{-\beta
 V(y)} \biggl( {1\over 2 \pi \beta} \biggr)^{{n\over 2}}
 \exp\!\bigg[-\frac{1}{2\beta} |sy-y|^2\bigg] \\ 
 + \frac{1}{L}
 \sum_{k=1}^{L-1} \int\!\rmd^n y \,\rmd^n x\, \rme^{-\beta V(y)}
 \biggl( {L\over 2 \pi \beta k} \biggr)^{{n\over 2}}
 \exp\!\bigg[-\frac{L}{2\beta k} |sx-y|^2\bigg] \\ 
 \times \biggl({L\over 2 \pi \beta (L-k)} \biggr)^{{n\over 2}}
 \exp\!\bigg[-\frac{L}{2\beta (L-k)} |y-x|^2\bigg].
\end{eqnarray*}
But $s$ is an orthogonal transformation and thus
$|sx-y|^2=|x-\trp{s}y|^2$.  Substituting this into the previous
formula and computing the resulting Gaussian integral over $x$ will
then show that, in fact, all the terms in the sum are equal and the
result is precisely the middle term in equation (\ref{e:lemeq}).  This
proves the first inequality.  The second inequality is obvious from
the definition of $\Zcl$, equation (\ref{e:defZcl}).
\end{proof}

Let us begin the proof of the theorem by defining for all $\re z >0$
\[
g(z) = \Tr(\opm \rme^{-z \ham}).
\]
It has been proven in \cite{jl:trform} that for the potentials we are
interested in the operator $\rme^{-z \ham}$ is trace-class on the
right half-plane.  Since $\opm$ is bounded, the trace converges and
can be given in terms of the eigenvectors $\psi_j$ and the
corresponding eigenvalues $E_j$ of the Hamiltonian  as the (numerable)
sum
\[
g(z) = \sum_{j} \langle{\psi_j}|{\opm\psi_j}\rangle \rme^{-z E_j}.
\]
Suppose next that $\re z \ge c >0$. Since then
$|\langle{\psi_j}|{\opm\psi_j}\rangle \rme^{-z E_j}|\le \rme^{-c E_j}$
and $\sum_j \rme^{-c E_j} = \Tr \rme^{-c \ham} < \infty$, we can apply
proposition \ref{th:anint} and conclude that $g$ is analytic on the
right half-plane and that
\begin{equation}\label{e:hampow}
g^{(k)}(z) = \sum_{j} \langle{\psi_j}|{\opm\psi_j}\rangle  (-E_j)^k
 \rme^{-z E_j} =  (-1)^k \Tr\!\left(\opm\ham^k\e^{-z\ham}\right).
\end{equation}

The operator defining the lattice trace, $\ope{T}(z)$, is for all 
$\re z > 0$ a Hilbert-Schmidt operator with a kernel
\[
K^{(T)}(a,b) = \frac{1}{(2 \pi z)^{n/2}} 
\exp\!\left(-\frac{1}{2 z}|a-b|^2 - z V(b)\right).
\]
Therefore, $\opm\ope{T}(z)$ is also Hilbert-Schmidt and it has a kernel
\[
\frac{1}{N!} \sum_{s\in S_N} (\pm )^s K^{(T)}(sa,b).
\]
Thus for any $L\ge 2$ the function 
$f_L(z)\equiv\Tr\!\left(\opm\ope{T}(z/L)^L\right)$ 
is well-defined and it has an integral representation
\begin{equation}\label{e:deffL} \fl
f_L(z) =   \frac{1}{N!} \sum_{s\in S_N} (\pm )^s \int\!\rmd^{nL} x\,
  \biggl( {L\over 2 \pi z} \biggr)^{{L n\over 2}} 
  \exp\!\bigg[-\frac{1}{z} P_L(sx(L),x) - z V_L(x)\bigg].
\end{equation}
We shall use equation (\ref{e:deffL}) to define also $f_1(z)$.

Suppose next that $\re z \ge c > 0$.  Then the integrand in
(\ref{e:deffL}) is bounded by  
$\left( {L/ 2 \pi c} \right)^{{L n\over 2}} \exp(- c V_L(x))$ 
which is integrable by assumption.  Thus the
conditions for proposition \ref{th:anint} are satisfied and each
$f_L(z)$ is an analytic function on the right half plane and all of
its derivatives can be computed by a differentiation inside the
integral.

We shall now first finish proving the theorem for the $k=0$ case, from
which the $k>0$ case follows quite easily. The uniform boundedness on
compact subsets will follow from the bound
\begin{equation}\label{e:flbound}
|f_L(z)| \le \rme^{-b_L \re z} \Zcl( a_L \re z)\mbox{, where }  
a_L =
 \cases{ 1 & for $L$ even \\  \case{L-1}{L} & for $L$ odd \\ }
\end{equation}
and $b_L = (L\,{\rm mod}\,2)V_{\rm min}/L$ with 
$V_{\rm min}=\inf_x V(x)$.  
Note that for even $L$ we have the simple bound 
$|f_L(z)|\le\Zcl(\re z)$.

By Lemma IV.1 of \cite{jl:trform}, for any bounded operator 
$\A$ and Hilbert-Schmidt operator $\T$,  
$|\Tr(\A\T^{2 \ell})|\le \norm{\A}\Tr(\T^\dagger\T)^\ell$.  
Since $\norm{\opm}=1$ and
$\norm{\T(z)}\le\rme^{-\re z V_{\rm min} }$, we thus have
\[
|f_L(z)| \le \rme^{-b_L \re z} \Tr( \T(a_L \re z/\ell)^\ell ) 
\mbox{, where } \ell = \left\lfloor L/2\right\rfloor.
\]
Choosing $s={\rm id}$ in the left hand side of equation
(\ref{e:deffL}) yields an integral representation for the trace in the
above equation. By lemma \ref{th:ikbound}, then 
$\Tr(\T(\beta/\ell)^\ell )\le \Zcl(\beta)$ 
and we have finished proving the
bound \ref{e:flbound}.

We next need the result that for all $\beta>0$,
\[
\lim_{L\to\infty} f_L(\beta)  = 
  \Tr \left(\opm \rme^{-\beta\ham}\right).
\]
The proof of this result can be done essentially identically to the
proof of Theorem III.4 in \cite{jl:trform} which contains the above
result for the case $\ope{P}=\ope{1}$.  We do not reproduce the proof
here.

We have now shown that the sequence $f_L$ consists of functions
analytic on the right half-plane, which converge on the real axis and
which are uniformly bounded on every compact subset of the half-plane.
By the Vitali convergence theorem, the sequence then converges on the
whole half-plane, the convergence is uniform on compact subsets and
the limit function is analytic.  Since the limit function is equal to
$g$ on the real axis, and $g$ is analytic,  it follows that the limit
function is given by $g$ on the whole half-plane.  Also, since the
convergence is uniform on compact subsets, then all  derivatives
converge as expected, $f_L^{(k)}(z) \to g^{(k)}(z)$.   By equation
(\ref{e:hampow}) we have now completed the proof of the first half of
theorem \ref{th:latconv}.

Suppose then that $0<c\le c'$ and $z$ satisfies $c\le \re z \le c'$.
By the Cauchy estimates for derivatives, then for all $0<t<1$,
\[
|f_L^{(k)}(z)| \le \frac{k!}{t^k c^k}  
  \sup_{(1-t) c\le \re w \le c'+t c} |f_L(w)|
\]
and it is thus enough to prove the uniform boundedness of $f_L(z)$.
However, since $\Zcl$ is now a  continuous function, this is an
obvious consequence of the inequality (\ref{e:flbound}).  Especially,
if $V$ is positive then $\Zcl$ is decreasing and we have for all even
$L$, $0<t<1$ and $\re z \ge c$,
\[
|f_{L}^{(k)}(z)| \le \frac{k!}{t^k c^k} \Zcl((1-t)c).
\]

\section{Simplification of the permutation sum}\label{sec:perms}

Let $r$ be a permutation and make a change of variables from $x$ to
$y$ to the lattice integral (\ref{e:trppm}) as defined by $x(k)=r
y(k)$ for all $k=1,\ldots,L$.   The Jacobian of this transformation is
one and, by assumption, $V_L(x)=V_L(y)$. The transformation $r$ is
orthogonal, and thus $|x(k)-x(k-1)|^2= |y(k)-y(k-1)|^2$ for all $k$
with the exception of $k=1$ for which we have
$|x(1)-sx(L)|^2=|y(1)-\trp{r}s r y(L)|^2$.  Therefore, the lattice
kernel satisfies $K_L(sx(L),x;z) = K_L( (r^{-1}s r)y(L), y;z)$.

Since $r$ was arbitrary, we can now conclude that the result of the
lattice integral depends only on the conjugate class of the
permutation.  We shall thus need a way of parameterizing the conjugate
classes and of choosing an easy representative element from each
class.  The solution is well-known and we present the results as they
are given in \cite{hamermesh}.

Every permutation can be decomposed into independent cyclic
permutations, cycles, and its conjugate class is determined by the
number $\nu_l$ of cycles of length $l$.  Conversely, a collection of
numbers $\nu_l$, $l=1,\ldots,N$,  defines a conjugate class provided
it satisfies the consistency condition $\sum_{l=1}^N l \nu_l = N$.
For each class, we can then choose a representative element as the
permutation which performs the cyclic permutations for consecutive
elements and puts the longest cycles first, e.g.\ if $N=7$ and
$\nu_3=1$, $\nu_2=1$, $\nu_1=2$, the representative permutation would
be $(1,2,3,4,5,6,7) \to (3,1,2,5,4,6,7)$.  The number of distinct
permutations in each class can be computed from $\nu_l$ by the formula
\[
\frac{N!}{\prod_{l=1}^N (l^{\nu_l} \nu_l !)} = \prod_{l=1}^N
 \bigg[\binom{N-\sum_{j=1}^{l-1} j \nu_j}{l\nu_l}
 \frac{(l\nu_l)!}{l^{\nu_l} \nu_l !}\bigg].
\]
Each cycle with an even number of elements has an odd parity and a
cycle with an odd number of elements has an even parity.  Thus the
parity of the permutations in the class is given by the formula
\begin{equation}\label{e:classparity}
(-1)^{\sum_{l=1}^N  (l-1) \nu_l} = (-1)^{N-\sum_{l=1}^N  \nu_l}.
\end{equation}

For our purposes, there exists a better alternative parameterization.
From $\nu_l$ define the new parameters $\lambda_l$, $l=1,\ldots,N$, via
\[
\lambda_l = \sum_{j=l}^N \nu_j.
\]
The inverse relation is clearly given by  
$\nu_l = \lambda_l-\lambda_{l+1}$  for $l=1,\ldots,N$, 
where it is understood
that $\lambda_l=0$ whenever  $l>N$.  The consistency condition for
$\lambda_l$ is now simply
\[
\sum_{l=1}^N\lambda_l=N\mbox{ and } \lambda_1\ge \lambda_2 \ge \cdots
 \ge \lambda_N \ge 0
\] 
and thus the generation of the conjugate classes reduces to the
generation of the partitions of $N$ into $N$ ordered non-negative
integers.  After this is done, we can compute $\nu_l$ and choose the
representative permutation, denoted from now on by $s_{(\lambda)}$, as
was explained in the previous paragraphs.  In addition, the number of
permutations in a class is given by
\begin{equation}\label{e:defcnu}
c_{(\lambda)} \equiv \frac{N!}{\prod_{l=1}^N
 [l^{\lambda_l-\lambda_{l+1}} (\lambda_l-\lambda_{l+1})!]}
\end{equation}
and, by (\ref{e:classparity}), the parity of the class is given by
$(-1)^{N-\lambda_1}$.

\section*{References}

%\bibliographystyle{\utildir{iopstyle}}
%\bibliography{\utildir{myabbr},\utildir{mrabbrev},\utildir{allrefs}}
%\end{document}

\end{document}